\documentclass[prc,aps,floatfix,nofootinbib,twocolumn,letterpaper]{revtex4} 
\usepackage{graphicx} 
\usepackage{amssymb} 
\usepackage{amsmath} 
\usepackage{amsfonts} 

\usepackage{color}

\renewcommand{\vec}[1]{\mbox{\boldmath $#1$}}

\def\be{\begin{equation}} 
\def\ee{\end{equation}} 
\def\im{{\rm Im}}
\def\re{{\rm Re}}

\def\imw{\im(w_{kk})}

\begin{document} 

\title{
Porter-Thomas fluctuations in complex quantum systems}

\author{K. Hagino}
\affiliation{ 
Department of Physics, Kyoto University, Kyoto 606-8502,  Japan} 

\author{G.F. Bertsch}
\affiliation{ 
Department of Physics and Institute for Nuclear Theory, Box 351560, 
University of Washington, Seattle, Washington 98915, USA}

\begin{abstract}
The Gaussian Orthogonal Ensemble (GOE) of random matrices has been widely employed to 
describe diverse phenomena in 
strongly coupled quantum systems.  An important prediction is that the decay rates of 
the GOE eigenstates fluctuate according to the distribution for one degree of
freedom, as derived by Brink and by Porter and Thomas.  However, we
find that the coupling to the decay channels can change the effective number 
of degrees of freedom from $\nu = 1$ to $\nu = 2$.  Our conclusions  are
based on a configuration-interaction Hamiltonian  originally 
constructed to test the validity of transition-state
theory, also known as Rice-Ramsperger-Kassel-Marcus (RRKM) 
theory in chemistry. 
The internal Hamiltonian
consists of two sets of GOE reservoirs connected by an internal channel.
We find that the effective number of degrees of freedom $\nu$ can vary from one to two
depending on the control parameter $\rho \Gamma$, where $\rho$ is the level
density in the first reservoir and $\Gamma$ is the level decay width.  
The $\nu = 2$ distribution is a well-known
property of the Gaussian Unitary ensemble (GUE);  our model demonstrates
that the GUE fluctuations can be present under much milder conditions. 
Our treatment of the  model permits an analytic derivation
for $\rho\Gamma \gtrsim 1$.  
\end{abstract}
 
\maketitle 

{\it Introduction.}
Random matrix theory was proposed by Wigner \cite{wi55} and extended by
Dyson \cite{dy62} to model generic features of complex quantum systems. 
The main idea is to consider an ensemble of Hamiltonians whose matrix elements 
are randomly generated to model the statistical properties
of the systems.
The theory has been widely employed to discuss properties in a variety of 
systems \cite{gu98} including nuclear spectra
\cite{ze96,WM09}, 
atomic spectra  \cite{na18}, electrons in mesoscopic
systems \cite{be97,al00}, unimolecular chemical reactions \cite{mi98}, 
quantum chromodynamics \cite{ve00} and
microwave cavity resonances \cite{alt1998,di15}.  
See also Ref. \cite{ji21} for a recent development of random state
technology, in which properties of random states are exploited 
to carry out numerical simulations for many-body
systems.

Prominent in the random matrix theory 
is the Gaussian Orthogonal Ensemble (GOE) which is used to simulate
Hamiltonians that obey 
time-reversal symmetry. It is well known 
that the eigenvalues and 
the eigenfunctions of GOE follow 
Wigner's semi-circular distribution for the average level density and Dyson's metrics for 
level spacings. Also, the wave-function amplitudes in the
GOE follow a Gaussian distribution.  This leads to a distribution of
decay widths that follow the $\nu=1$ Porter-Thomas (PT) distribution \cite{th56}
\be
P_\nu(x) = \frac{\nu}{2x_0\Gamma(\nu/2)}\left(\frac{\nu x}{2x_0}\right)^{\nu/2-1}\,
e^{-\nu x/2x_0}
\label{PT}
\ee
where $x_0$ is the mean value of the widths. 
This distribution and the one for $\nu = 2$ were
 originally proposed by Brink \cite{br55}.
The index $\nu$ has the values 
$\nu=1$ for the GOE and $\nu = 2$ for the Gaussian Unitary Ensemble (GUE) composed
of complex Hermitian matrices. Since the Hamiltonian matrices governing the quantum systems 
are often real, it is commonly  assumed that the distribution of decay rates  
can be derived from the GOE.

In reality, the $\nu=1$ PT distribution can be violated for several 
reasons, most obviously when the Hamiltonian violates time-reversal
invariance as in electron dynamics in a magnetic field. 
In nuclear physics, the $\nu =1$ distribution has recently 
become controversial \cite{koehler2010,koehler2011,shriner2000} and 
other mechanisms have been suggested to explain deviations
\cite{volya2015,bogomolny2017, 
CAIZ2011, 
volya2011,weidenmuller2010,fy15}. 

In this paper, we revisit this problem using a random matrix model 
we developed in Ref. \cite{verA}.
The model was
constructed   to assess the validity of  
transition 
state theory\cite{BW39,tgk96,thh83,MJ94,M74,LK83,MR51,M52}. 
The internal states of the system are represented by two GOE Hamiltonians connecting with each other via 
bridge states. Each GOE Hamiltonian is augmented by an imaginary energy
$-i\Gamma/2$ on the diagonal associated with direct decays from the states.
Hamiltonians based on two interacting GOE reservoirs have been studied
previously \cite{harney1986,alt1998}, but limited to purely real Hamiltonians. 
In our reaction model, the Hamiltonian also contains an explicit
entrance channel that is coupled to the first GOE reservoir.
Those reservoir states can decay directly or pass to the
second reservoir through the bridge channel.  
We will show below that the decay rate from the second GOE Hamiltonian follows 
the $\nu=1$ distribution when $\Gamma_a$ for the first GOE matrix is small, 
changing gradually to the $\nu=2$ distribution as $\Gamma_a$ increases. 
Note that the internal Hamiltonian is real, but becomes effectively complex due
to the boundary conditions imposed by the coupling to the entrance and decay
channels.
 
{\it Model.}
The Hamiltonian in our model is a matrix acting on states in a
discrete-basis representation.  The bridge channel consists of two states that are connected to
each other and to the sets of GOE reservoir states.  The Hamiltonian is defined as 
\be
\label{H} 
{\mathbf H} = \left[\begin{matrix} 0 & t_1 & 0 & 0 & 0 & 0\cr
t_1 & 0 &  {\vec v}_2^T&0&0&0 \cr
0 &  {\vec v}_2 & H^{\rm goe}_a - i\Gamma_a/2 &  {\vec v}_3 & 0 & 0 \cr
0&0& {\vec v}_3^T &0 &t_2 &0 \cr
0&0&0  & t_2 & 0 & {\vec v}_4^T\cr
0&0&0&0&\vec v_4 & H^{\rm goe}_b-i\Gamma_b/2\cr
\end{matrix}\right]. 
\ee
The first two entries in the vector space are associated with states in the
entrance channel; the parameter $t_1$ is a hopping matrix element connecting
adjacent states in the channel.  The entries in
the fourth and  
fifth rows and columns apply to the bridge states. The third and sixth rows
and columns represent $N_g\times N_g$ subblocks containing the GOE
Hamiltonians with $g=a$ or $b$.
The matrix elements in the $H^{\rm goe}_g$ submatrices are taken from the 
GOE ensemble \cite{WM09},
\be
\langle i|H^{\rm goe}_g|j \rangle 
=\langle j|H^{\rm goe}_g|i \rangle = r_{ij}  v_g (1 + \delta_{i,j} )^{1/2}. 
\ee
Here $r_{ij}$ is a random number from a Gaussian distribution
of unit dispersion, $\langle r_{ij}^2 \rangle = 1$, and $v_g$ is the 
root-mean-square value of the matrix elements. 
The vectors $\vec v_k$ connect the channels to the GOE states, and 
we assume that their matrix elements are given  as 
$\vec v_k(i) = r_i v_k$, where $r_i$ is random 
with $\langle r_i^2 \rangle = 1$ 
and 
$v_k$ is an overall scaling
factor.  It will be convenient to parameterize the
derived analytic formulas in terms of the GOE level density 
$\rho_{0g} = N_g^{1/2}/\pi v_g$ at the center of the spectrum and the limiting eigenvalues 
$E_{mg} = \pm 2
N_g^{1/2} v_g$.

As described in 
Ref. \cite{verA} and in the Supplementary Material, 
the GOE states can be treated 
implicitly in a reduced Hamiltonian, leaving only the four channel
amplitudes explicit:
\be
\label{H2} 
H_{\rm red}=
\left[\begin{matrix}
0 & t_1 & 0  & 0 \cr 
t_1 &w_{22} &  w_{23} & 0 \cr
0 & w_{23} &  w_{33} & t_2 \cr
0 & 0 & t_2 & w_{44} \cr
\end{matrix}\right].
\ee
Here the $w_{kk'}$ are self-energies associated with the states in the
channels.  They are
given by
\be
\label{wkk'-def}
w_{kk'} =  \vec v_k \cdot (E - H_g^{\rm goe}+i \Gamma_g/2)^{-1} \cdot \vec
v_{k'}
\ee
where $E$ is the total energy of the reaction.
These are evaluated with $(H_g^{\rm goe},\Gamma_g)=(H_a^{\rm goe},\Gamma_a)$ 
for $w_{22}, w_{23}$, and $w_{33}$, and with 
$(H_g^{\rm goe},\Gamma_g)=(H_b^{\rm goe},\Gamma_b)$ for $w_{44}$.
Since the spectrum of $H_g^{\rm goe}$ is purely real,  
the inverse matrix expression (\ref{wkk'-def}) always exists. 
The reaction cross section $\sigma_{k\ell}$ associated with an entrance channel $k$ leading
to an exit channel $\ell$  can be computed as a kinematic cross section for
channel $k$ multiplied by 
a transmission factor $T_{k\ell}$,
\be
\sigma_{k\ell} = \sigma_k T_{k\ell}. 
\ee  
Our model has only one entrance channel and we drop the index $k$ in
the formulas below.  There are many exit channels associated with the imaginary decay
widths; we add together all the contributions passing through states in
reservoir $a$ to define $T_a$ and similarly for reservoir $b$.
The total inelastic transmission factor $T$ is then given by
$T = T_a + T_b$.  
Notice that $T$ and $T_b$ are proportional to 
$\Phi_{12}$ and $\Phi_{34}$, respectively, where $\Phi_{ij}$ expresses 
the probability flux from channel site $i$ to $j$. 
Formulas for $T$ and $T_a$ expressing their
dependence on the Hamiltonian parameters are derived in the Supplementary
Material. 

A particularly interesting physical observable is 
the probability $P_b$ of a reaction whose decay products out of the $b$
reservoir in competition with other decay modes,
\be
P_b  = \frac{T_b}{T}. 
\ee 
This is closely related to the branching ratio $B_r = T_b/T_a$ discussed in
Ref. \cite{verA}. 
As derived in the Supplementary Material, 
$P_b$  can be
expressed in terms of the Hamiltonian parameters as
\be
\label{Pb_full}
P_b = 
  \frac{t_2^2 |w_{23}|^2 {\rm Im}(w_{44})}
{{\rm Im}(w_{22}) |s|^2  -
\im(w_{23}^2 w_{44} s^*)},
\ee
where $s = w_{33} w_{44} - t_2^2$.

{\it Fluctuation statistics.} 
We derived the transition-state formula in 
Ref. \cite{verA} 
by estimating
the mean value of $B_r$ from the statistical properties of the
self-energies.  For that estimate we evaluated the expectation values of 
the diagonal self-energies and their off-diagonal squares
$|w_{23}|^2$ and $w^2_{23}$. The results are shown in Table
\ref{statistical}, together with additional statistical properties
needed in the present context.  See Refs. \cite{lo00,fe20} and the Supplementary Material for
their derivation.
\begin{table*}[htb] 
\begin{center} 
\begin{tabular}{|c|cc|cc|} 
\hline 
$x$  & $\langle \re \,x \rangle$ & $\langle \im \,x \rangle$ & SD($\re \,x$) &
SD($\im \,x$) \\
\hline
&&&&\\
$w_{kk}$  & 0 &  $-\pi v_k^2\rho_{0g}$  &   
$\left(\frac{2\pi v_k^4\rho_{0g}}{\Gamma_g}\right)^{1/2}$ &
$\left(\frac{2\pi v_k^4\rho_{0g}}{\Gamma_g}\right)^{1/2}$ \\
&&&&\\
$w_{kk'}$& 0 &  0  &  
$\left(\frac{\pi v_k^2 v_{k'}^2\rho_{0g}}{\Gamma_g}\right)^{1/2}$  &
$\left(\frac{\pi v_k^2 v_{k'}^2\rho_{0g}}{\Gamma_g}\right)^{1/2}$  \\
&&&&\\
$|w_{kk'}|^2$ & 
$\frac{2\pi v_k^2 v_{k'}^2\rho_{0g}}{\Gamma_g}$  &&  
 $\frac{2\pi v_k^2 v_{k'}^2\rho_{0g}}{\Gamma_g}$ &   \\
&&&&\\
$w_{kk'}^2$ &  $-\frac{\pi v_k^2 v_{k'}^2\rho_{0g}}{E_{mg}}$  & 0 & 
$\frac{2\pi v_k^2 v_{k'}^2\rho_{0g}}{\Gamma_{g}}$   &
$\frac{2\pi v_k^2 v_{k'}^2\rho_{0g}}{\Gamma_{g}}$   \\
\hline 
\end{tabular} 
\caption{
Expectation values and standard deviations  SD$(x) = \sqrt{\langle x^2\rangle
-\langle x\rangle^2}$ of self-energy expressions appearing in Eq.
(\ref{Pb_full}).  The statistical properties have been evaluated at $E=0$ in the limits of
large $N_g$ and $ (\rho_{0g})^{-1} \ll \Gamma_g \ll E_{mg} $ .
It is assumed that $k \ne k'$ in the entries with
subscript $kk'$. 
\label{statistical}
} 
\end{center} 
\end{table*} 

In assessing how the statistical properties of the self-energies affect
$P_b$, 
we first note that $w_{23}^2$ 
is small compared to the other terms in the 
denominator of Eq. (\ref{Pb_full}).  This is due to its inverse dependence on
$E_{mg}$, since that energy
is large compared to all other energy scales. Also, the fluctuation in
the diagonal self-energy can be neglected for large GOE spaces since
it varies as $N_g^{-1/4}$ times its expectation value.
Thus, the entire fluctuation in $P_b$ 
can be attributed to its dependence on $|w_{23}|^2$ in the numerator.  From Table I we
see that 
its standard deviation is equal to its expectation value.   
In the Porter-Thomas family of distributions (\ref{PT}), the 
$\nu=1$ standard deviation is twice its
expectation value while the $\nu=2$ distribution is equal to 
the expectation value.  One can also infer that the fluctuations in 
$w_{kk'}$ have two independent degrees of freedom by noting that the
cross-correlator $\langle(\re\,w_{kk'})(\im\,w_{kk'})\rangle$ vanishes in
the limit considered above. Thus the real and imaginary parts can be
considered separate degrees of freedom\footnote{We are indebted to Y.
Alhassid for pointing out this connection.}.  
This is our analytic evidence that
the fluctuations in transition-state theory follow the $\nu=2$
distribution in the overlapping resonance region, $\rho_{0a}\Gamma_a \gtrsim 1$.

For the remainder of the article we explore numerically the distribution
for a range of $\rho_{0a}\Gamma_a$ extending well into the isolated resonance
region\footnote{The Green's function for the isolated resonance region
has also been studied analytically \cite{ro04}.}, $\rho_{0a}\Gamma_a \ll
1$. 
Fig. 1 shows 
the distribution of $P_b$ for the Hamiltonian parameters given in the
caption.  One can see that the
numerically sampled distribution agrees well with the Porter-Thomas
for $\nu=2$ degrees of freedom. 

\begin{figure}[tb] 
\includegraphics[width=0.8\columnwidth]{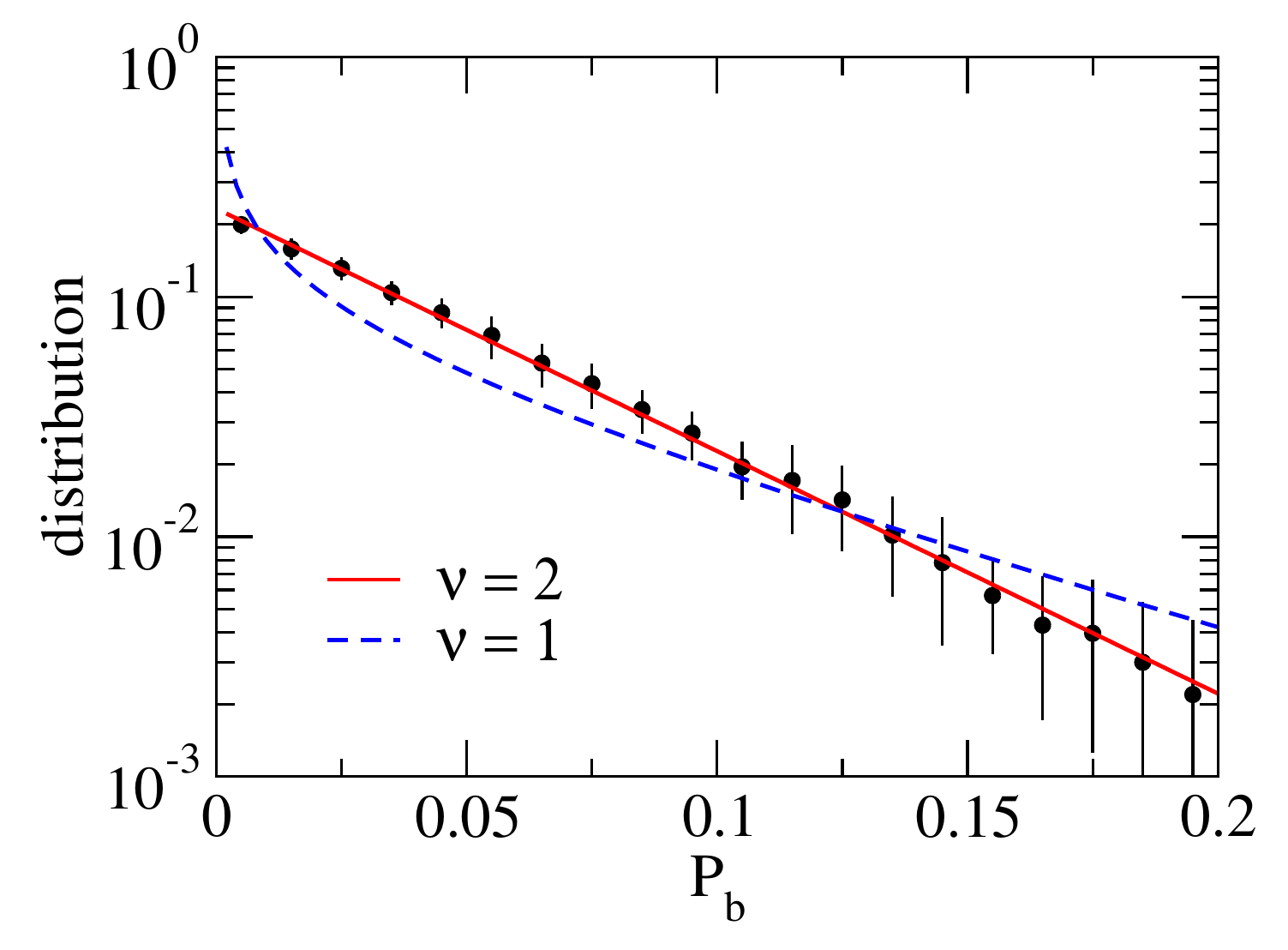}         
\caption{Distribution of numerically sampled decay probabilities $P_b$ (black circles) 
compared with the Porter-Thomas distributions for $\nu = 1$ (dashed line)
and $\nu = 2$ (solid line). The dimension of the two GOE spaces are
$N_g=100$ and their Hamiltonian parameters $v_g,v_k,v_{k'},\Gamma_g$ are set
to 0.1.  The hopping matrix elements in the channel spaces are taken as $t_i
= 1$.  The mean values and the 
root-mean-square (rms) 
deviations for the numerical sampling are
calculated for 50 histogrammed runs, each 
of which is constructed for 500 samples. 
\label{pb-fig} 
}
\end{figure} 
To understand the deviation from the $\nu=1$ 
Porter-Thomas distribution, Fig. 2 shows 
the distribution of the probability $P_b$ 
for several values of $\Gamma_a$, setting $t_2 = -(10 \Gamma_a)^{1/2}$
and keeping the other parameters the same as in Fig. 1.  We
wish to keep the expectation value $\langle P_b \rangle$ constant
as $\Gamma_a$ is varied.  This is achieved in
the transition-state formula  Eq. (38) of 
Ref. \cite{verA} 
by changing
$t_2$ as described.   The two curves in each panel show the
fits to the PT distribution with $\nu=1$  and $\nu=2$.
When $\Gamma_a$ is much smaller than 
$v_g$ and $\Gamma_b$, as in Fig. 2(a), the distribution is consistent with the 
$\nu=1$ PT distribution.
As $\Gamma_a$ increases, it 
gradually deviates from that, and eventually comes close to the $\nu=2$
PT distribution. 
We have checked that the distribution
is insensitive to the decay widths in the second reservoir over a
broad range of the parameter
$\rho_{0b}\Gamma_b$.

We also carried out a least-squared fit of $\nu$ in the PT distribution to the histogrammed data 
with results shown in Fig. 3.  It comes out close to $\nu=1$ for small
control parameter $\rho_{0a}\Gamma_a$
and to $\nu = 2$ for moderate and large $\rho_{0a}\Gamma_a$.  
We have also plotted on the Figure the function
$\nu(y) = (1+8.28y^2)/(1+ 3.81y^2)$ with $y = \rho_{0a}\Gamma_a$ as a purely
phenomenological description of fitted $\nu$ parameters. 

\begin{figure}[tb] 
\includegraphics[width=\columnwidth]{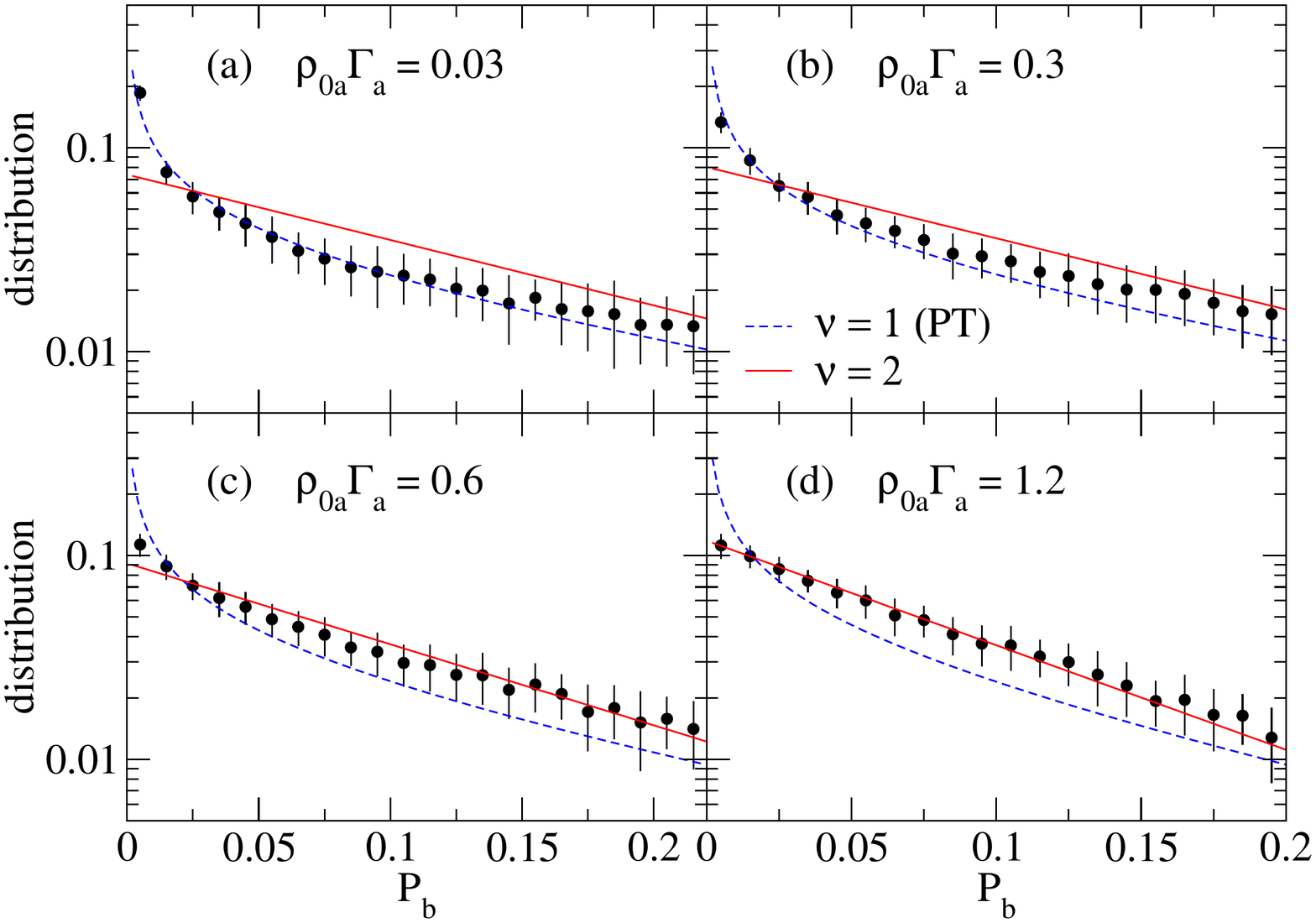}  
\caption{ 
The distribution of the transmission probability for the
second reservoir, $P_b$, for several values of $\Gamma_a$ and
$t_2=-(10 \Gamma_a)^{1/2}$ as 
explained in the text.
The dots with error bars were calculated with 50
histogrammed samplings as in Fig. 1. The dashed and the solid
curves denote the $\nu=1$ and 2 PT distributions, respectively.
}
\end{figure} 

\begin{figure}[tb] 
\includegraphics[width=0.8\columnwidth]{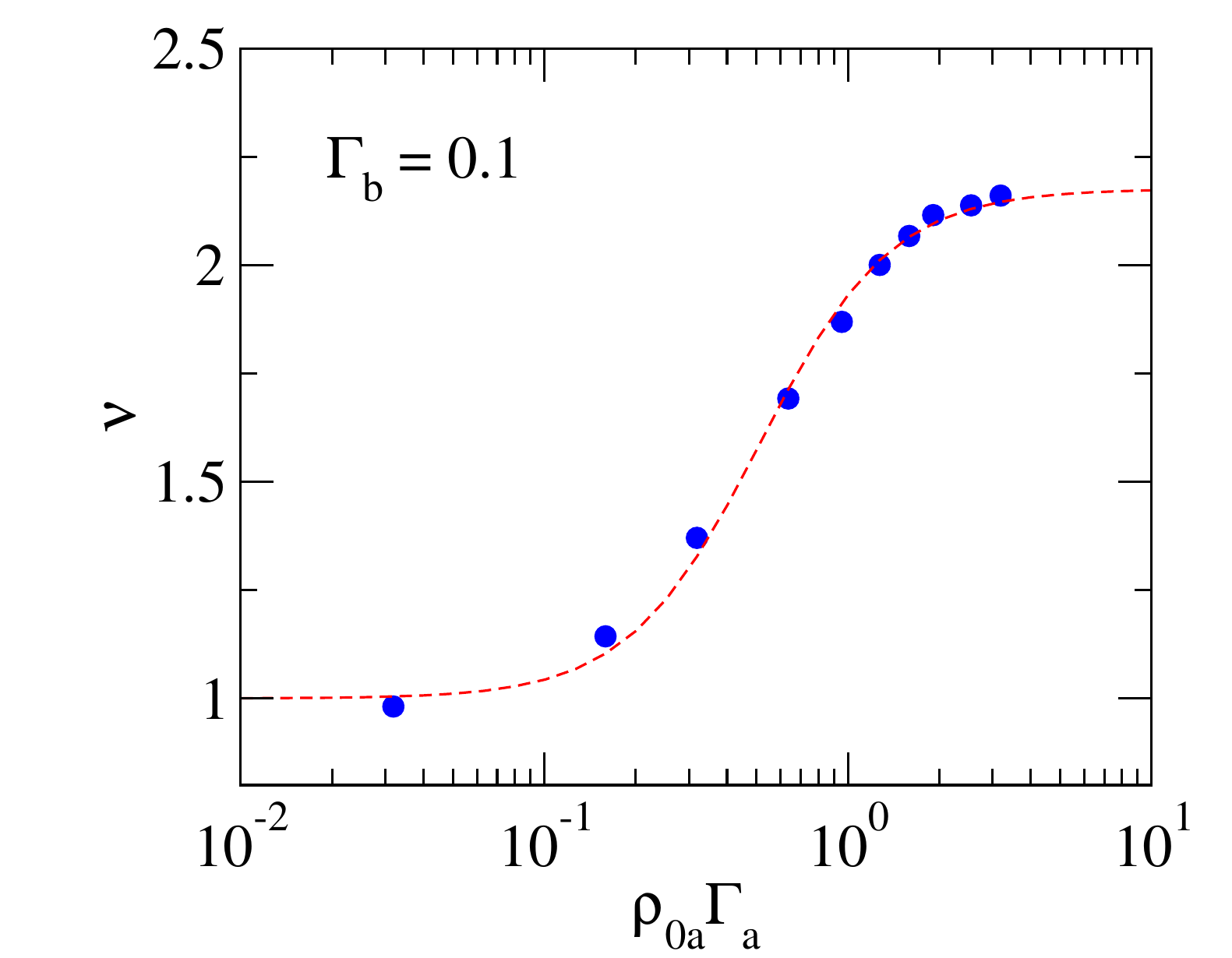} 
\caption{Fitted values to the number of degrees of freedom $\nu$ as a
function of the control parameter $\Gamma_a \rho_{0a}$.  The Hamiltonians
are defined in the same way as in Fig. 2.
The dashed line shows an empirical fit, 
$\nu(y) = (1+8.28y^2)/(1+ 3.81y^2)$ with $y = \rho_{0a}\Gamma_a$.
Note that the $\nu$ exceeds 2 in the asymptotic region $\rho_{0a}\Gamma_a
\gg 1$.  This may be a finite size effect, but we have not examined this
possibility.  }
\end{figure} 

{\it Summary.}
Making use of random matrixx theory, we have applied a Hamiltonian 
to fluctuations in reactions of complex quantum systems. 
The model had been
previously proposed to find the limits of validity of the transition-state
theory of averaged reaction quantities.  
It is common wisdom that fluctuations in decay rates
associated with a transition state  in a time-reversal-invariant
Hamiltonian follow the PT distribution for one degree of
freedom.  However, the effective Hamiltonian is complex 
when boundary conditions arising from
other channels are taken into account.  When those decay widths
are comparable or larger than the average level spacing, the
fluctuations approach the PT distribution for $\nu$ = 2.  In the
model, the key quantity responsible for fluctuations is the 
quantity $w_{23}$ which depends on
Green's function for the Hamiltonian of the first reservoir.  
For real Green's functions the fluctuations are also
real, corresponding to a single degree of freedom.  However, if
the states in the reservoir can decay directly into continuum
channels, the Green's function is complex and the fluctuations
approach those of a complex quantity with independent variations
in the real and imaginary parts.  This behavior leads to reaction
rates that follow the $\nu=2$ PT distribution.  The model shows
the crossover from one distribution to the other, with the
control parameter identified as $\rho_{0a}\Gamma_a$.  

A crossover from $\nu=1$ to $\nu=2$ has also been studied in 
random matrix models \cite{al98,al00}, interpolating between the
GOE and the GUE ensembles.  However, it is not
clear from such studies  how to relate the complex  matrix
elements to physical quantities when the underlying Hamiltonian
is purely real.     

The present model might be useful in the methodology for determining  the
effective number of channels in transition-state theory.  In Ref.
\cite{po90} the effective number of channels in a unimolecular reaction
was estimated from
a formula based on the $\nu=1$ PT distribution \cite{mi90},
\be
\nu_{\rm eff}  = 2\langle \Gamma
\rangle^2/(\langle\Gamma^2\rangle-\langle\Gamma\rangle^2). 
\ee   
The authors found that their theoretical calculations were a factor
of two off.  Depending on the direct decay widths of the initial molecule,
the explanation might be the factor of two difference from the
$\nu = 1$ and $\nu = 2 $ PT distributions.

Previously, it has been shown in nuclear physics that a coupling to continuum state 
narrows the distribution, leading to $\nu$ which is smaller than one if the 
distribution is fitted with Eq. (\ref{PT}) \cite{CAIZ2011}. 
This was not realized in our model, as the value of $\nu$ was found to be 
between 1 and 2. 
In any case, it would be interesting if the deviation from the Porter-Thomas 
distribution discussed in that paper could be observed experimentally. 

\bigskip

We thank Yoram Alhassid for linking the $\nu=2$ distribution to
the fluctuations of
$w_{23}$ in the complex plane.  We also thank him, Yan Fyodorov, 
and Hans Weidenm\"uller 
for additional comments on the manuscript.
This work was supported in part by
JSPS KAKENHI Grant Numbers JP19K03861 and JP21H00120.

\section{Supplementary Material}

\subsection{Decay probability $P_b$}

For completeness and to make the paper self-contained, 
we here provide a short derivation of the effective Hamiltonian (4). 
Call the vector of states that the Hamiltonian acts on
\be
\boldsymbol\Psi = (\phi_1,\phi_2,\Psi_a,\phi_3,\phi_4,\Psi_b). 
\ee
The amplitudes $\phi_1,\phi_2$ are the nearest ones in the entrance
channel and $\phi_3,\phi_3$ are in the bridge channel.  $\Psi_a$ and 
$\Psi_b$ are sets of amplitudes for states in the GOE reservoirs.
For a fixed
amplitude $\phi_1$ the Hamiltonian equation
\begin{equation}
({\mathbf H}-E)
\left(\begin{matrix} \phi_1 \\ \phi_2 \\ \Psi_a \\ \phi_3 \\ \phi_4 \\ \Psi_b 
\end{matrix}\right)= \left(\begin{matrix} 0\\0\\0\\0\\0\\0\\
\end{matrix}\right) 
\end{equation}
can be solved for the remaining amplitudes by simple matrix operations.
This is carried out in two steps.  In the first step the amplitudes $\Psi_a$
are expressed in terms of $\phi_2$ and $\phi_3$.  Similarly the amplitudes
in $\Psi_b$ are expressed in terms of $\phi_4$.  This reduces the
Hamiltonian equation to the form 
\be
\label{H2eq} 
\left[\begin{matrix} 
w_{22} - E &  w_{23} & 0 \cr
w_{23} &  w_{33}-E & t_2 \cr
0 & t_2 & w_{44}-E \cr
\end{matrix}\right]
\left(\begin{matrix} \phi_2 \cr \phi_3 \cr \phi_4 \cr
\end{matrix}\right) =
-\left(\begin{matrix} t_1\phi_1 \cr 0 \cr 0 \cr
\end{matrix}\right).
\ee

We next derive formulas for the decay probabilities following the lines
presented in Ref. \cite{verA}. 
For simplicity we restrict the energy to $E=0$, 
which is in the middle of the spectrum distributions of the 
GOE matrices in the Hamiltonian (\ref{H}).  Eq. (\ref{H2eq})
is easily solved for amplitudes $\phi_2,\phi_3$ and $\phi_4$ in terms
of $\phi_1$; the solution is
\begin{align}
\phi_2 &= -(w_{33} w_{44} -t_2^2) t_1 \phi_1 /D \\
\phi_3 &= w_{23} w_{44} t_1 \phi_1 / D\\
\phi_4 &= -w_{23} t_2 t_1 \phi_1 / D
\end{align}
where
\be
D = w_{22} w_{33}w_{44} - w_{22} t_2^2 - w_{23}^2 w_{44}.
\ee

The probability we seek can be expressed in terms of the probabilities
fluxes $\Phi_{ij}$ from channel site $i$ to neighboring channel site $j$ as
\be
P_b = \frac{\Phi_{34}}{\Phi_{12}}.
\ee
The individual fluxes are computed by the standard quantum relation 
\be
\Phi_{ij} = \langle i | H | j \rangle 
\im (\phi_i \phi_j^*),
\ee
where 
$\langle i | H | j \rangle$ 
is the hopping matrix element between the two sites, 
that is, 
$\langle 1 | H | 2 \rangle=t_1$ and 
$\langle 3 | H | 4 \rangle=t_2$.
The results are
\be
\Phi_{12} = -2t_1^2\left[{{\rm Im}(w_{22}) |s|^2  -
\im(w_{23}^2 w_{44} s^*)}\right]\frac{1}{|D|^2} |\phi_1|^2 
\ee
where $s = w_{33}w_{44} - t^2_2$ and
\be
\Phi_{34} =  -2 t_1^2 t_2^2 |w_{23}|^2 \im (w_{44})\frac{1}{|D|^2} |\phi_1|^2. 
\ee
The transmission factors are easily expressed in terms of
the fluxes; for our purposes here we only need the ratio
of the two fluxes given in Eq. (\ref{Pb_full}). 

\subsection{Variances of self-energy quantities}
Statistical properties of the GOE Green's function have been
derived in Refs. \cite{lo00} and \cite[App. C]{fe20} for the
limits given in Table I.  We follow the same method here to
determine the quantities needed in Eq. (\ref{Pb_full}).
The derivations are based on an eigenfunction representation of
the self-energies,
\begin{equation}
w_{kk'}
=\sum_{j}^{N_g}\frac{\langle \vec{v}_k|\phi_{j}\rangle\langle\phi_{j}|
\vec{v}_{k'}\rangle}{E-E_{j}+i\Gamma_g/2}
\label{self-energy}
\end{equation}
where $E_{j}$ and $\phi_{j}$ are the eigenvalues and eigenfunctions
of a GOE Hamiltonian of dimension $N_g$.
The overlap $\langle\vec{v}_k|\phi_{j}\rangle$ is given by
$ \langle\vec{v}_k|\phi_{j}\rangle = v_k r_{j}$,  where $v_k$ is a
Hamiltonian parameter and $r_{j}$ is a Gaussian variable satisfying
$ \langle r_{j} r_{j'} \rangle = \delta_{jj'}$.
The $r$ variables
associated with different $\vec v_k$ vectors are
distinguished as $r$ and $r'$.  They satisfy $\langle r_{j} r'_{j'} 
\rangle = 0$.

We first consider an ensemble average of the 
diagonal self-energy, $w_{kk}$, at $E=0$. 
Since the eigenvector components are
uncorrelated with each other or with the eigenenergies, 
the ensemble average can be expressed as 
\begin{equation}
\langle w_{kk}\rangle 
= \langle (v_kr_{j})^2\rangle
\left\langle\sum_{j}^{Ng} \frac{1}{-\tilde{E}_{j}}\right\rangle
= v_k^2
\left\langle\sum_{j}^{Ng} \frac{1}{-\tilde{E}_{j}}\right\rangle
\end{equation}
where $\tilde E_{j} = E_{j} - i\Gamma_g/2$. 
By replacing the sum over $j$ by the energy integral with the 
level density $\rho(E)=\rho_0\sqrt{1-(E/E_{mg})^2}$, 
one obtains \cite{verA} 
\begin{equation}
\langle w_{kk}\rangle 
\approx v_k^2 \rho_0 {\cal I}_1(i\Gamma_g/2 E_{mg})
\approx -i \pi v_k^2 \rho_0\,\,\,{\rm for}\,\,\, \Gamma_g \ll E_{mg}, 
\end{equation}
where 
\be
\label{stieltjes2}
{\cal I}_1(z) =  \int^{+1}_{-1} dx \frac{\sqrt{1-x^2}}{z-x} = 
\pi\left(z- \sqrt{z+1}\sqrt{z-1}\right). 
\ee
taking the principal value of the square root. 

To derive the formula for the variance of $\imw$, 
we start with the equation for its second moment,
\begin{eqnarray}
&&\langle [\imw]^2\rangle   \nonumber \\
&&= \left<\sum_{i i'}^{N_g} 
(\vec v_k\cdot \vec \phi_i)^2(\vec v_k\cdot \vec \phi_{i'})^2
{\rm Im}
\left(\frac{1}{\tilde{E}_i}\right)
{\rm Im}
\left(\frac{1}{\tilde{E}_{i'}}\right)
\right>.  \nonumber \\
\end{eqnarray} 
The numerator factors are $v_k^4 r_i^2 r_j^2$ in the new notation.  The
expectation value of the product of variables is
\begin{align}
\langle r^2_i r^2_j\rangle &= \langle r^2\rangle^2 (1 -\delta_{ij})
+ \langle r^4\rangle \delta_{ij} \\
&= 1 + 2 \delta_{ij}.
\end{align}  
Neglecting the fluctuation in the $E_i$, the expectation value
becomes
\be
\langle [\imw]^2\rangle = v_k^4\left( \im \sum_i \frac{1}
{\tilde{E}_i}\right)^2 + 2v_k^4\,\sum_i \left({\rm Im}
\frac{1}{\tilde{E}_i}\right)^2.
\ee
The first term is just the square of $\langle\imw\rangle$ and the 
second term is the variance.

As before, we replace the sums by an integral.  In dimensionless form the
required integral is ${\cal I}_2$ given by
\begin{align}
{\cal I}_2(y) = \int_{-1}^{+1} dx \frac{(1-x^2)^{1/2}}{(x^2 + y^2)^2}
   =\frac{\pi }{2 y^3 \sqrt{1+y^2}}\approx \frac{\pi}{2 y^3}. 
\end{align}
Evaluating the integral
at $y = \Gamma_g/2 E_m$, one obtains for the variance
\be
\label{wkk-var}
\langle [\imw]^2\rangle -  \langle \imw\rangle^2 = 2 \pi v_k^4 \rho_0/\Gamma_g.
\ee 

The variance for the real part of $w_{kk}$ can be evaluated in a similar way. 
Using the integral 
\begin{equation}
{\cal I}_3(y) = \int_{-1}^{+1} dx \frac{x^2(1-x^2)^{1/2}}{(x^2 + y^2)^2}\approx \pi/2y, 
\end{equation}
one obtains 
\be
\langle [{\rm Re}(w_{kk})]^2\rangle -  \langle 
{\rm Re}(w_{kk})\rangle^2 = 2 \pi v_k^4 \rho_0/\Gamma_g, 
\label{wkk-var2}
\ee 
which coincides with the variance of the imaginary part of $w_{kk}$. 

Let us next consider the square of the absolute value of 
off-diagonal self-energies $w_{kk'}$ at $E=0$ with $k\neq k'$. 
To determine the expectation value of this quantity, we first express it as 
\be
\label{wkkpasq}
\langle|w_{kk'}|^2\rangle  
= v_k^2v_{k'}^2\langle r_ir_jr_i'r_j'\rangle
\left\langle
\sum_{i,i'}^{N_g} \frac{1}
{\tilde{E}_i\tilde{E}_{i'}^*}  \right\rangle.
\ee 
Using 
\begin{equation}
\langle r_ir_jr_i'r_j'\rangle=\langle r_ir_j\rangle^2=\delta_{i,j},
\end{equation}
one obtains 
\begin{equation}
\langle|w_{kk'}|^2\rangle =  v_k^2 v_{k'}^2 \frac{\rho_0}{E_m} {\cal I}_4(c) 
\approx  2\pi v_k^2v_{k'}^2 \frac{\rho_0}{\Gamma_g}, 
\end{equation}
where $c = \Gamma_g/2E_m$, and ${\cal I}_4$ is given by 
\begin{equation}
{\cal I}_4(y) = \int_{-1}^{+1} dx \frac{(1-x^2)^{1/2}}{(x^2 + y^2)}
=\frac{\pi}{y}\left(\sqrt{1+y^2}-y\right)\approx \frac{\pi}{y}.
\end{equation}
The separate variance of the real and imaginary parts of
$w_{kk'}$ can be evaluated in the same way using integrals
${\cal I}_4$ and ${\cal I}_4-y^2{\cal I}_2$ (see Eq. (\ref{wkkp-var2}) below).
Note that the integral required for the 
correlation $\langle (\re\,w_{kk'})(\im\,w_{kk'})\rangle$ vanishes identically.

The variance of $|w_{23}|^2$ is given by a product of four sums over
eigenstates with  8 Gaussian variables in the numerator.  The expectation
value of the product is 
\begin{eqnarray}
&&\langle r_i r_j r_k r_l r'_i r'_j r'_k r'_l\rangle =
\langle r_i r_j r_k r_l\rangle^2 \\
&&=
\langle r^2\rangle^4
(\delta_{ij}\delta_{kl} + \delta_{ik}\delta_{jl} + \delta_{il}\delta_{jk})
(1 - \delta_{ijkl}) \nonumber \\
&&~~~~+ \langle r^4 \rangle^2 \delta_{ijkl})\\
&&=
(\delta_{ij}\delta_{kl} + \delta_{ik}\delta_{jl} + \delta_{il}\delta_{jk})
 + 6  \delta_{ijkl}. 
\label{r8}
\end{eqnarray}
Here  $\delta_{ijkl} = \delta_{ij}\delta_{ik}\delta_{il}$. We next insert
these restrictions into the sums over eigenstates.  Two of the first
three terms in Eq. (\ref{r8}) reduce the sums  to $\langle\im(w_{kk'})\rangle^2$. The
third term gives $\langle w_{kk'}\rangle^2$ which we have seen can be
neglected for physical parameter sets. For parameters sets such that the last
term is also small, the variance is
\be
\label{wkkp-var}
\langle|w_{kk'}|^4\rangle - \langle|w_{kk'}|^2\rangle^2 =
\langle|w_{kk'}|^2\rangle^2.
\ee

In a similar way, one finds 
\begin{eqnarray}
&&\langle{\rm Re}(w_{kk'})^2\rangle - \langle{\rm Re}(w_{kk'})\rangle^2 
\nonumber \\
&&=
\langle{\rm Im}(w_{kk'})^2\rangle - \langle{\rm Im}(w_{kk'})\rangle^2 
=
\pi\rho_0\frac{v_k^2v_{k'}^2}{\Gamma_g}
\label{wkkp-var2}
\end{eqnarray}
and
\begin{eqnarray}
&&\langle[{\rm Re}(w_{kk'}^2)]^2\rangle - \langle{\rm Re}(w_{kk'}^2)\rangle^2 
\nonumber \\
&&=
\langle[{\rm Im}(w_{kk'}^2)]^2\rangle - \langle{\rm Im}(w_{kk'}^2)\rangle^2 
=
4\left(\frac{\rho_0\pi}{\Gamma_g}\right)^2
v_k^4v_{k'}^4. \nonumber \\
\end{eqnarray}
To check these estimates, we have compared them with a numerical sampling
of the ensembles.  The results for a few parameter sets are shown in 
Table \ref{wkk-table}; the agreement is quite satisfactory.

Notice that the integrals ${\cal I}_1$, ${\cal I}_2$, ${\cal I}_3$, and 
${\cal I}_4$ correspond to the integrals $I_1$, $I_4$, $I_5$, and $I_3$ in 
Ref. \cite{verA}, respectively.

\begin{table*}[htb] 
\begin{center} 
\begin{tabular}{|c|c|c|c|} 
\hline 
\hline 
$N_g$ & Type &  sampled   &  analytic   \\
\hline
100& $w_{kk}$ &   $-(0.02\pm 0.48) -(1.02\pm0.47)i$ & $(0\pm 0.45)-(1\pm 0.45) i$
\\
&$|w_{kk'}|^2$ & $0.23 \pm 0.27$  &  $0.2\pm0.2$ \\ 
\hline 
400 &$w_{kk}$ &   $-(0.00025\pm 0.61) -(1.95\pm0.55)i$ & $(0\pm0.63)-(2\pm0.63)i$  
  \\
&$|w_{kk'}|^2$ & $0.41 \pm 0.41$  &  $0.4\pm0.4$ \\
\hline 
900 &$w_{kk}$ &   $-(0.083\pm 0.80) -(3.06\pm0.77)i$ & $(0\pm 0.77)-(3\pm0.77)i$
\\
&$|w_{kk'}|^2$ & $0.60 \pm 0.66$  &  $0.6\pm0.6$ \\
\hline 
1600 &$w_{kk}$ &   $-(0.030\pm 0.91) -(4.06\pm0.96)i$ & $(0\pm0.89)-(4\pm0.89)i$
\\
&$|w_{kk'}|^2$ & $0.83 \pm 0.96$  &  $0.8\pm0.8$ \\
\hline 
\hline 
\end{tabular} 
\caption{
Expectation values and statistical fluctuations of diagonal and off-diagonal 
self-energies associated
with the coupling of channels to GOE ensembles.  
The GOE Hamiltonian parameters $v_g,v_k,v_{k'},\Gamma_g$ are set
to 0.1 and the hopping matrix elements $t_i$ are set to one.
The entries labeled
``analytic" were calculated with these parameters in the statistical
formulas (\ref{wkk-var},\ref{wkk-var2},\ref{wkkp-var}). The entries labeled ``sampled" were
obtained with 100 numerically calculated samples. 
\label{wkk-table}
} 
\end{center} 
\end{table*}

\end{document}